\documentclass[journal=jacsat,manuscript=article]{achemso}

\usepackage[version=3]{mhchem}

\usepackage[colorlinks,linkcolor={blue},citecolor={blue},urlcolor={blue}]{hyperref}
\usepackage{xcolor}
\usepackage{bm}
\usepackage{amssymb}
\pdfoutput=1

\author{Micheal S. Dimitriyev}
\email{msdim@tamu.edu}
\affiliation[TAMU]
{Department of Materials Science \& Engineering, Texas A \& M University, College Station, Texas}
\alsoaffiliation[UMASS]
{Department of Polymer Science \& Engineering, University of Massachusetts, Amherst, MA}
\author{Benjamin R. Greenvall}
\affiliation[UMASS]
{Department of Polymer Science \& Engineering, University of Massachusetts, Amherst, MA}
\author{Rejoy Matthew}
\affiliation[UMASS]
{Department of Polymer Science \& Engineering, University of Massachusetts, Amherst, MA}
\author{Gregory M. Grason}
\email{grason@umass.edu}
\affiliation[UMASS]
{Department of Polymer Science \& Engineering, University of Massachusetts, Amherst, MA}

\title[Not even metastable]
  {Not even metastable: Cubic double-diamond in diblock copolymer melts}

\abbreviations{IR,NMR,UV}

\begin{document}

\begin{tocentry}





\includegraphics[width=\textwidth]{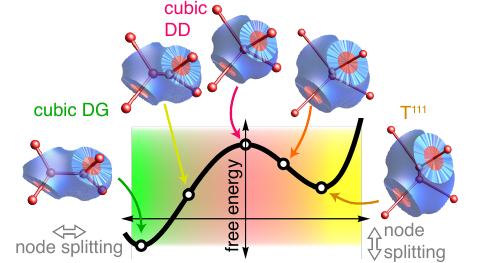}

\end{tocentry}

\begin{abstract}
We study the thermodynamics of continuous transformations between two canonical, cubic network phases of block copolymer melts:  double-gyroid, an equilibrium morphology for many systems; and double-diamond, often thought to be a close competitor.  We use a strong-segregation approach to compute the free energy of double network morphologies as a function of two structural parameters that convert between two limiting cubic cases; a tetragonal stretch of the unit cell in combination with fusion of pairs of trihedal gyroid nodes into tetrahedral diamond nodes.  For the simplest case of conformationally symmetric diblock melts, we find that cubic double-diamond sits at an unstable saddle point that is continuously deformable into the lower free energy gyroid, as well as a second metastable, tetragonal network composed by trihedral nodes.  We confirm the broad instability of double-diamond at finite segregation using self-consistent field studies and further show that it derives directly from the entropic free energy cost of chain packing in the tubular domains of tetrahedral nodes. Correspondingly, we demonstrate two factors that quench the entropic cost of packing in the tubular domain -- homopolymer blending and elastic asymmetry between the blocks -- promoting double-diamond to a metastable state by way of free energy barrier that separates it from double-gyroid.
\end{abstract}

\section{Introduction}

Triply-periodic, polycontinuous networks remain among the most fascinating and functionally useful morphologies formed by block copolymers (BCPs) and other amphiphilic assemblies~\cite{Hyde1997_ch4, Lee2014}. 
The cubic double-gyroid (DG) and double-diamond (DD) phases are archetypal forms, which in simple AB-type systems are composed of two inter-connected tubular network subdomains of one composition (e.g.~A) separated by the hyperbolically-shaped slab-like subdomain of the other (e.g.~B)~\cite{Thomas1986,Hajduk1994,Reddy2021}.  
Gyroid and diamond morphologies differ in network connectivity, composed respectively of inter-catenated loops with 10 or 6 nodal junctions, with each node connected to either 3 or 4 neighbors by tubular struts~\cite{Charvolin1987}. 
While many amphiphilic systems (e.g.~lyotropic membranes) exhibit both stable gyroid and diamond structures~\cite{HydeAndersson1984, Squires2000, Oka2015}, it is widely-established based on extensive self-consistent field theory (SCFT) studies in BCP melt systems that the DG is thermodynamically favored over DD, under nearly all conditions where equilibrium network phase are found~\cite{Matsen1994,Matsen1996,Reddy2022,Dimitriyev2024}.  
Nevertheless, DD morphologies have been observed in experiments on linear diblock melts~\cite{Chang2021,Feng2023,Shan2024}, suggesting it to be at least a metastable competitor of DG~\cite{Chang2024}.  
Additionally, DD is predicted to become an equilibrium morphology under more rarefied conditions, such as in binary or tertiary blends~\cite{Matsen1995,Martinez-Veracoechea2009b,Lai2021,Xie2023}, consistent with its experimental observations~\cite{Winey1992,Takagi2017,Takagi2019,Takagi2021}.  

Our present study is motivated by the recent experimental observation of coexistence and transformation between these two cubic network archetypes within the same BCP sample~\cite{Shan2024}.  
Direct 3D tomographic imaging of morphologies in polystyrene-b-polydimethylsiloxane (PS-PDMS) copolymers reveals regions of both DG and DD within the same sample, separated by a wide inter-phase boundary surface, remarkably showing that the transformation between the limiting DD and DG morphologies takes place {\it continuously} and without breakage of either of the (PDMS) network domains.
Consequently, there is a process by which loops of trivalent nodal regions, dubbed as ``mesoatoms,''~\cite{GrasonThomas2023} interconvert with loops of tetravalent nodal mesoatoms, with consequent chain re-arrangements.
The observation of a direct and continuous transition between DG and DD raises basic questions about the structural and thermodynamic pathway connecting these cubic cousins.  
Chief among these questions are: What are the optimal pathways for, and consequences of, changing network connectivity? What determines the free energy barrier separating DG and DD, if it exists? 

In this Letter, we use a combination of strong-segregation theory (SST) and SCFT to study uniform, continuous transformations between DG and DD.  
Motivated by a continuous, non-intersecting transformation between G and D minimal surfaces~\cite{FogdenHyde1999, Schroder-Turk2006} that has been applied to explain observations in lyotropic amphiphiles~\cite{Squires2005, Oka2015}, we consider a \textit{tetragonal} transformation pathway shown schematically in Fig.~\ref{fig:1}.
This pathway is described by two parameters, measuring respectively, tetragonal stretch of the initially cubic repeat of gyroid, and fusion of pairs of gyroid nodes perpendicular to the stretch direction into tetravalent diamond nodes.  
We exploit a novel implementation of {\it medial SST}~\cite{Reddy2021} to compute the free energy landscape as a function of these two parameters. 
Surprisingly, this approach reveals that there is no barrier between DD and DG for the simplest case of linear AB diblocks melts.  
In contradiction to the intuition that DD is a metastable competitor to DG, we find it lies at a saddle point and is unstable to both cubic DG as well as the recently predicted tetragonal T$^{111}$ double network phase~\cite{Chen2023}.
We use SCFT to show that the lack of stability for DD extends to finite-segregation, and then further investigate the conditions under which DD becomes at least metastable, giving examples of blends with (minority type) homopolymer and large conformational asymmetry in linear diblocks.

\section{Geometry of tetragonal DG-to-DD intermediates }

\begin{figure}[t!]
    \centering
    \includegraphics[width=6.5in]{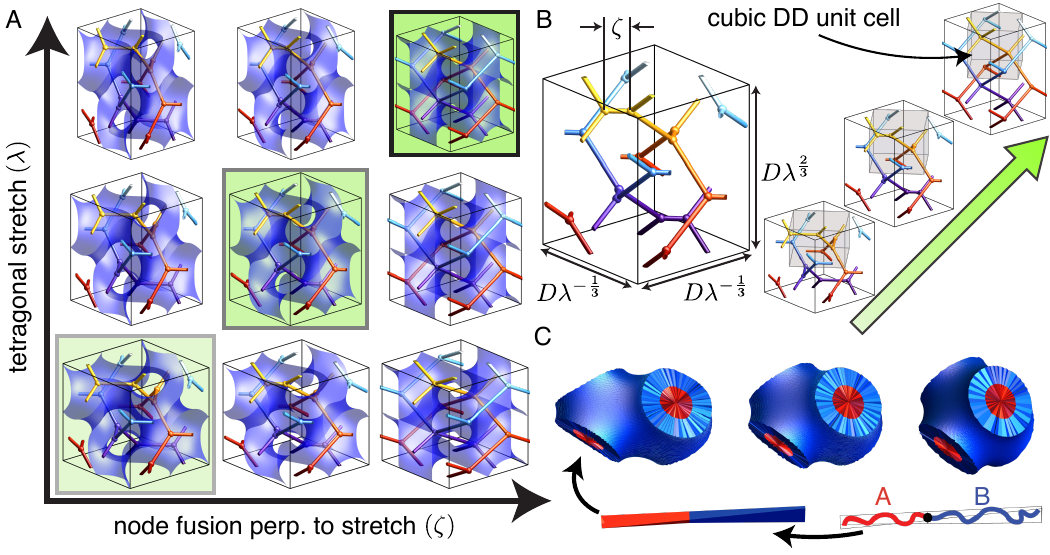}
    \caption{\label{fig:1} Tetragonal DG/DD transformation pathway model. (A) The two-parameter family of skeletal graphs forms a landscape of candidate approximations to triply-periodic minimal surfaces, including highlighted the gyroid, diamond, and intermediate tetragonal structures. (B) The gyroid-diamond path combines a tetragonal stretch of the unit cell ($\lambda$) and non-affine fusion ($\zeta$) of nodes along the struts lying perpendicular to the stretch direction. (C) AB diblock copolymers occupy small volumes that tessellate the volumes around pairs of nodes that fuse along the double gyroid/double diamond pathway.
    }
\end{figure}

Instead of using minimal surface geometry~\cite{FogdenHyde1999}, we parametrize the continuous tetragonal transformation between DG and DD by the pair of skeletal graphs that thread through tubular subdomains of the double network, with structural degrees of freedom encoded in nodal positions, Fig.~\ref{fig:1}.
Cubic DG and cubic DD are described by the (10,3)-a and (6,4) nets with 3- and 4-valent nodal junctions, respectively~\cite{Wells}, and can be interconverted continuously via a combination of tetragonal deformation and node fusion.
The affine, volume-preserving, tetragonal deformation of the unit cell is described by $\lambda$, which maps points via $(x,y,z)\mapsto (x,y,\lambda z)/\lambda^{1/3}$, chosen such that $\lambda = 1$ is the cubic DG reference state.
A second parameter, $\zeta_\perp$, describes the degree of non-affine splitting of pairs of trivalent nodes in planes perpendicular to the tetragonal stretch ($z$) axis, such that a pair of strut-sharing nodes with the same $z$-coordinate $\mathbf{r}_1(1)$ and $\mathbf{r}_2(1)$ are moved along the strut via
\begin{equation}
{\bf r}_i(\zeta_\perp) = \zeta_\perp {\bf r}_i(1) + (1 - \zeta_\perp)\frac{{\bf r}_1(1)+{\bf r}_2(1)}{2} \, .
\end{equation}
Decreasing $\zeta_\perp$ from 1 shrinks the length of 4 of the initial 12 edge directions of the (10,3)-a net, such that as $\zeta_\perp \to 0 $ the neighbor pairs are fused into tetravalent node at the center of the initial edge.
As illustrated in Fig.~\ref{fig:1}(A), at the point $(\zeta_\perp, \lambda) = (0,\sqrt{2})$, the initial cubic cell of gyroid $(\zeta_\perp, \lambda) = (1,1)$ is distorted into 4 copies of the (primitive) cubic unit cell of diamond, rotated by $\pi/4$ relative to the initial cubic basis directions.
Notably, for intermediate values of these parameters this landscape represents a generalized class of deformed gyroid and diamond structures, including the recently-reported ``tetragonal gyroid''~\cite{Wang2025}. 


\section{Free energy landscape: strongly segregated, conformationally-symmetric diblocks}


To assess the thermodynamic cost of each candidate structure for neat AB diblock melts, we use medial strong-segregation theory (mSST)~\cite{Reddy2022,Dimitriyev2023} to determine the free energy per chain in the limit of infinite block segregation strength, $\chi N \to \infty$.  
Our approach exploits the nodal positions described above as collective coordinates to parameterize the space-filling arrangements of diblocks.
The resulting space-filling mesoatomic packings of strongly-segregated AB chain trajectories continuously distort in shape and position according to smooth changes in $(\zeta_\perp, \lambda)$, as detailed in SI section S1.   

\begin{figure}[t!]
    \centering
    \includegraphics[width=6.5in]{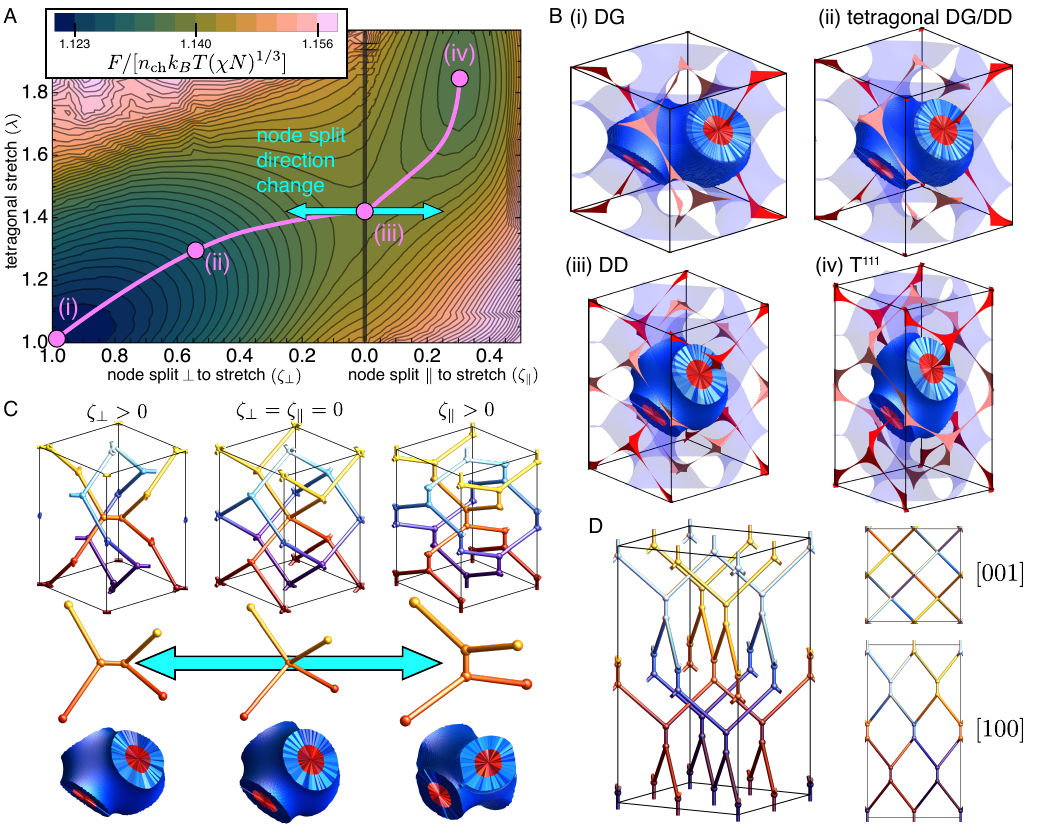}
    \caption{\label{fig:2} Free energy landscape of the DG/DD transformation, and beyond. (A) Free energy landscape in terms of the tetragonal stretch $\lambda$ and fusion parameters perpendicular to ($\zeta_\perp$) and along ($\zeta_\parallel$) the stretch direction. A continuous path passing passing through the DD saddle point and ending at cubic DG and the metastable T\textsuperscript{111} phase is shown in pink. (B) Mesoatom structures shown for (i) cubic DG, (ii) a tetragonal intermediate to DG/DD, (iii) cubic DD, and (iv) T\textsuperscript{111}. (C) Transition between transversely-split and longitudinally-split nodes at $\lambda = 1$, with focus on subgraphs and mesoatoms to highlight the T1 process. (D) Alternative choice of T\textsuperscript{111} unit cell shown with $[100]$ and $[001]$ projections.
    }
\end{figure}

In Fig.~\ref{fig:2}(C) we show the free energy per chain ($F/n_{\rm ch}$) over the $(\zeta_\perp, \lambda)$ landscape between cubic DG and DD for the conformationally-symmetric case of segment lengths $a_A =a_B$ at A-block fraction $f_A = 0.29$, where lamella and cylinder morphologies have (nearly) equal free energy and double networks compete for equilibrium.
We find that cubic DG at $(1,1)$, with $F_{\rm DG}/[n_{\rm ch} k_B T (\chi N)^{1/3}] \approx 1.120$, is the clear global minimum.  
Consistent with prior calculations that cubic DD has higher free energy per chain, the transition from $(1,1)$ to $(0,\sqrt{2})$ increases the free energy per chain to $F_{\rm DD}/[n_{\rm ch} k_B T (\chi N)^{1/3}] \approx 1.139$.  
Remarkably, however, cubic DD appears at an unstable saddle point, corresponding to continuous paths in the $(\zeta_\perp, \lambda)$ space that take DD to DG {\it without a free energy barrier}.  
While the structures at distinct $(\zeta_\perp, \lambda)$ points are free to equilibrate volume, a similar barrier-free instability exists in a fixed-volume landscape, shown in SI Fig.~S2.
This implies that for strongly-segregated diblock melts, cubic DD is in fact a transition state between DG and another triply periodic network phase, rather than the long-suspected metastable state.

We determine the second ``post-DD'' (meta)stable network structure by extending the 2-parameter tetragonal space past the point of node fusion as $\zeta_\perp \to 0$ by introducing a new parameter $\zeta_\parallel$ that measures node splitting {\it along} the tetragonal stretch axis.  
When the 4-valent nodes split into pairs of 3-valent nodes along the tetragonal axis this leads to a tetragonal, bicontinuous double network phase known as T\textsuperscript{111}, recently predicted as a low free energy SCFT candidate via machine learning~\cite{Chen2023} with trihedral mesoatom pairs shown in Fig.~\ref{fig:2}(D). 
We predict a local minimum in free energy per chain at relatively large tetragonal stretch $\lambda \approx 1.86$ and $\zeta_\parallel \approx 0.31$ with free energy \mbox{$F_{\rm T^{111}}/[n_{\rm ch} k_B T (\chi N)^{1/3}] \approx 1.135$}, indicating that T\textsuperscript{111} is indeed metastable to DG.
Therefore, the class of DG-like skeletal graphs, notated as (10,3)-a nets, where nodes fuse in the plane perpendicular to the tetragonal stretch axis (parameterized by $\zeta_\perp$), can be continuously deformed into the class of T\textsuperscript{111}-like skeletal graphs, notated as (10,3)-b nets~\cite{Wells}, where nodes fuse along the tetragonal stretch axis (parameterized by $\zeta_\parallel$).
This continuous ``flipping'' of the fused strut is analogous to the T1 process in foams and cellular geometries~\cite{Glazier1992}.
The fact that the tetravalent cubic DD is an unstable transition between two 3-valent networks suggests that trihedral node fusion is thermodynamically unstable for at least the prototypical case of conformationally-symmetric diblock melts.

\section{Instability of DD at finite segregation }

\begin{figure}[t!]
    \centering
    \includegraphics[width=6.5in]{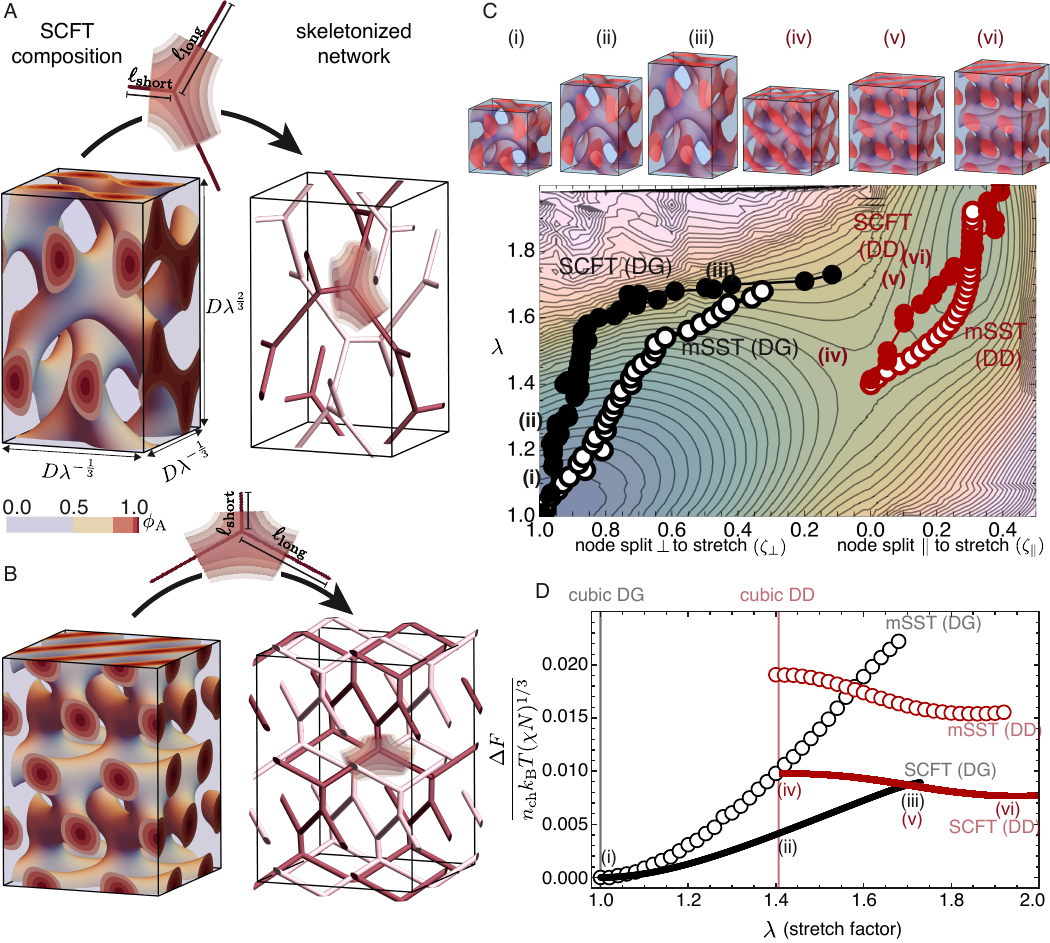}
    \caption{\label{fig:3} 
Deformation at finite segregation. (A) $\lambda = 1.67$ SCFT density field (left). Extracted skeletonized network (right); node splitting ($\ell_{\rm short}$) occurs $\perp$ to the stretch direction.
(B) $\lambda \simeq 1.92$ converged SCFT density field (left) and extracted skeleton (right), where node splitting occurs $\parallel$ to the stretch direction.
(C) Calculated node splitting ($\zeta_\perp$, $\zeta_\parallel$) for SCFT (solid markers) and mSST (open markers); black points correspond to deformations of cubic DG and red points correspond to deformations of cubic DD. The underlying free energy landscape is reproduced from Fig.~\ref{fig:2}(A) (mSST). Density fields along the DG (i-iii) and DD (iv - vi) show node fusion/fission perpendicular/parallel to the stretch direction, respectively. Snapshots are shown at ($\zeta_\perp = 1, \lambda = 1$), ($\zeta_\perp \simeq -0.92., \lambda \simeq \sqrt{2}$), ($\zeta_\perp = -0.70, \lambda = 1.67$), ($\zeta_\parallel = 0, \lambda = \sqrt{2}$), ($\zeta_\parallel \simeq 0.2, \lambda \simeq 1.70$), and ($\zeta_\parallel \simeq 0.35, \lambda \simeq 1.92$). Structures in (A) and (B) are the same as those shown in (iii) and (iv).
(D) Change in free energy per chain for increasing $\lambda$.}
\end{figure}

Predictions in Fig.~\ref{fig:3} show cubic DD is unstable as $\chi N \to \infty$, raising basic questions about the stability over a broader range of segregation conditions and diblock compositions, which we address here using self-consistent field theory (SCFT) calculations (SI section S2).  
Since it is not possible to directly control the internal network structure or displacement of mesoatom centers in SCFT, we instead imposed tetragonal transformations of the cubic cell parameters of DG and DD phases (parameterized by the tetragonal distortion $\lambda$ from cubic DG), and ${\it measure}$ the degree of inter-node fusion by skeletonizing the resulting A-block segment density fields (SI section S2.C) and extracting the $\zeta_{\perp,\parallel}$ as shown in Figs.~\ref{fig:3}(A) and (B).
We first consider a case of intermediate segregation strength, $\chi N = 25$, and A-block fraction $f_A = 0.33$, within the parameter range where cubic DG is favored thermodynamically.
This yields the $(\zeta,\lambda)$ paths shown in Fig.~\ref{fig:3}(C), which show remarkable agreement with corresponding minimal free energy inter-node fusion mSST results.  
While upon increasing $\lambda>1$ pairs of trivalent nodes indeed tend to fuse, the fixed tetragonal strain pathway starting from DG does not intersect with the cubic DD saddle point.
Rather, there is lower free energy branch of fixed tetragonal DG morphologies that extend up to tetragonal stretch values $\lambda \approx 1.7 > \sqrt{2}$, consistent with the shape of the mSST free energy landscape.

Imposing tetragonal stretch starting from cubic DD (i.e.~for $\lambda > \sqrt{2}$) leads to a distinct solution branch, involving fission of the 4-coordinated node along the stretch direction, as shown in Fig.~\ref{fig:3}(C), again evolving to the T\textsuperscript{111} network (Figs.~\ref{fig:3}(B) and (C.iv)) at $\lambda \simeq1.92$, which is only slightly larger than optimal value from mSST ($\lambda \simeq1.85$). 
In Fig.~\ref{fig:3}(D) we compare the free energies of fixed-$\lambda$ DG and DD branches from mSST to SCFT, showing strong qualitative agreement---despite the vast difference in segregation conditions---including definitive instability of cubic DD.

\begin{figure}[t!]
    \centering
    \includegraphics[width=6.5in]{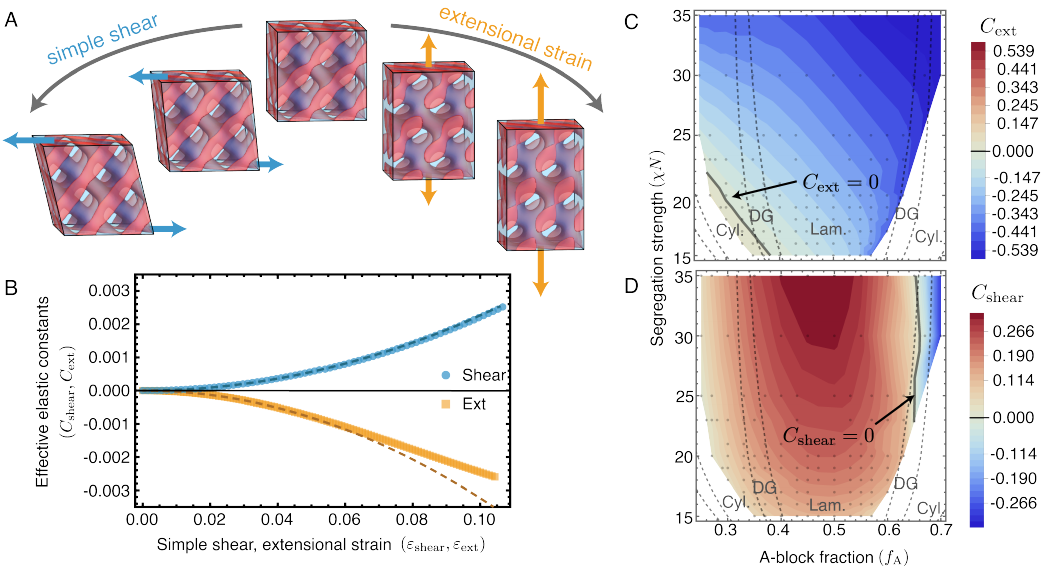}
    \caption{\label{fig:4}
    SCFT-computed deformation response of cubic DD. Simple shear and uniaxial extension, shown in (A), results in free energy per chain data (B), the concavity of which determines stability. The stability of DD as a function of $\chi N$ and A-block fraction $f_A$ are shown for uniaxial extension (C) and simple shear (D), superimposed on the diblock phase diagram from ref.~\cite{Matsen2012}. Red contours represent stable regions; blue contours are unstable. Scale of concavity is shown in the color bar, in units of $n_{\rm ch} k_B T$. Gray points correspond to SCFT data. 
    }
\end{figure}

We next probe the (in)stability of cubic DD over a much broader range of finite segregation and diblock composition by measuring the change in free energy to small distortions of the cubic unit cell.  
Small variations in lattice parameters can be expressed in terms of a symmetric strain tensor $\bm{\varepsilon}$, and to lowest order, the change in free energy $\Delta F$ of cubic DD with strain occurs at quadratic order in $\bm{\varepsilon}$, and can be expressed as (see SI)
\begin{equation}
    \frac{\Delta F}{n_{\rm ch}k_B T} = \frac{C_{\rm dil}}{2} \varepsilon_{\rm dil}^2+\frac{C_{\rm shear}}{2} \varepsilon_{\rm shear}^2+\frac{C_{\rm ext}}{2} \varepsilon_{\rm ext}^2
\end{equation}
where $\varepsilon_{\rm dil}$, $\varepsilon_{\rm shear}$ and $\varepsilon_{\rm ext}$ describe isotropic dilation, simple shear and extensional shear strains, as depicted in Fig.~\ref{fig:4}(A), and $C_{\rm dil}$, $C_{\rm shear}$ and $C_{\rm ext}$ are the corresponding effective elastic constants.  
We extract these coefficients from quadratic fits of the free energy computed at imposed unit cell parameters within SCFT.
For example, as shown Fig.~\ref{fig:4}(B), the deformation free energy of cubic DD at $\chi N = 30$ and $f_A = 0.35$ is stable to simple shear ($C_{\rm shear}>0$) but unstable with respect to extension ($C_{\rm ext}< 0$), corresponding to tetragonal distortions.
In Fig.~\ref{fig:4}(C,D), we plot the values of $C_{\rm ext}$ and $C_{\rm shear}$ for cubic DD over a broad range of $\chi N$ and $f_A$ for conformationally symmetric AB diblock melts.  
Indeed, $C_{\rm shear}>0$ almost everywhere, approaching instability only for very large values of $f_A$, in parameter ranges where ``inverse" morphologies form.  
In contrast, as shown in Fig.~\ref{fig:4}(C), the cubic DD is {\it almost always} unstable to extensional strain, over nearly the full range of equilibrium ordered phases, only becoming marginally metastable for especially low $f_A$, outside of window of equilibrium double network formation.  
Notably, cubic DG is shown to be broadly stable to both simple and extensional shear over similar parameter windows (Fig.~S5).  

\section{Conditions for DD metastabilty}


\begin{figure}
    \centering
    \includegraphics[width=6.5in]{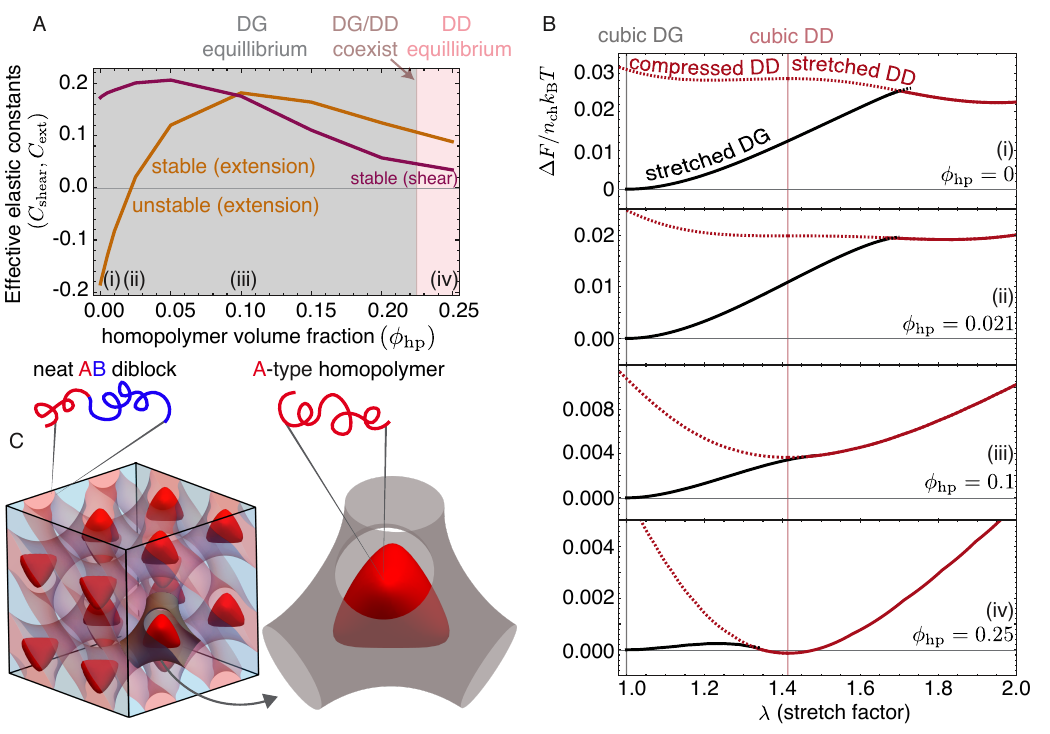}
    \caption{\label{fig:5}
    Stability of Double Diamond in minority homopolymer-doped blends. 
    (A) Effective modulus of Double Diamond to extension (orange) and shear (purple) with increasing homopolymer volume fraction. The single phase equilibrium morphology (gray, DG; red, DD) is shown underlaid and is separated by a small region of two-phase coexistence.  
    (B) Relative free energy change (per chain) with extension for fixed homopolymer doping. Deformations of a cubic Double Gyroid are shown in black and deformations of a cubic Double Diamond are shown in red; the optimal solution branch is shown as a solid trace.
    (C) $\phi_{\rm hp} = 0.10 $ Double Diamond expanded unit cell (left) and nodal ``mesoatom'' (right); the intermaterial dividing surface (IMDS, $\phi_{\rm B} = 0.5$) is shown in gray, while high concentrations of homopolymer ($\phi_{\rm hp} > 0.5$) are shown as solid red.
    }
\end{figure}

The rather generic instability of cubic DD for the simplest class of diblock melts in SCFT is surprising given observations of forming cubic DD under several experimental conditions, raising the question: Under what conditions is cubic DD at least metastable?  
First we consider blending a homopolymer additive that is miscible with the tubular network-forming block, which has been shown previously to promote the equilibrium selection of DD over DG~\cite{Matsen1995PRL, Matsen1995, Martinez-Veracoechea2009, Martinez-Veracoechea2009b, Takagi2017, Takagi2021}.
We consider the conditions of ref.~\cite{Martinez-Veracoechea2009b} where $f_A = 0.33$ diblocks were blended with A-type homopolymer with chain length $N_{\rm hp} = 0.8 N$ at $\chi N = 25$. 
Fig.~\ref{fig:5}(A) shows the effective elastic constants of cubic DD as a function of increasing homopolymer volume fraction $\phi_{\rm hp}$ computed via variable cell SCFT (see SI).  
While $C_{\rm ext}<0$ for pure diblocks, stability with respect to tetragonal deformations is achieved for homopolymer volume fractions as low as $\phi_{\rm hp} \gtrsim 3\%$, as shown in Fig.~\ref{fig:5}(B) by the local minimum at $\lambda = \sqrt{2}$ and is substantially less than the volume fraction $\phi_{\rm hp} \gtrsim 22.5\%$ needed to stabilize cubic DD over DG.
The localization of guest homopolymer at the centers of tetravalent DD mesoatoms, visualized in Fig.~\ref{fig:5}(C), is consistent with the long-standing interpretation that sequestering minority component ``guests" at the node centers of DD relaxes local hot-spots in the tubular domain chain packing, i.e.~regions of particularly high chain stretching free energy cost~\cite{Matsen1996, Dimitriyev2023}.  
We find that the same effect that promotes the equilibrium stability of DD also promotes its metastability, but at a much smaller fraction of guest molecules.

\begin{figure}[t!]
    \centering
    \includegraphics[width=6.5in]{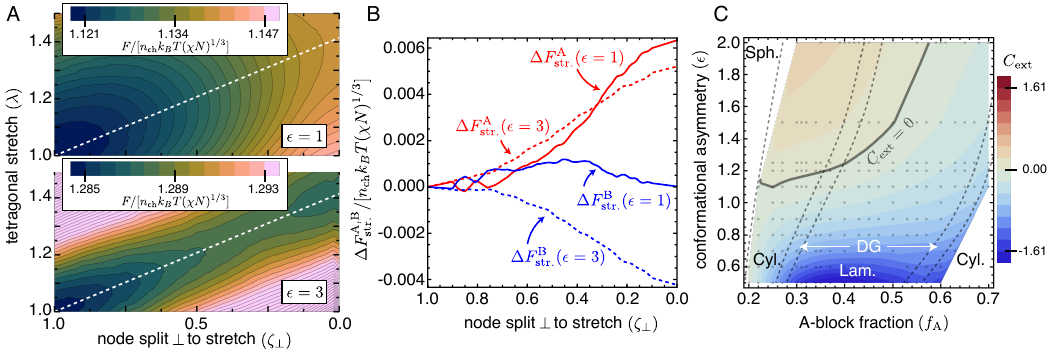}
    \caption{\label{fig:6} Stabilization of cubic DD using conformational asymmetry. (A) Free energy landscapes and pathways joining cubic DG and cubic DD for cases of conformational symmetry ($\epsilon = 1$, top) and asymmetry ($\epsilon = 3$, bottom). (B) Relative stretching free energy $\Delta F^{\rm A,B}_{\rm str.} = F^{\rm A,B}_{\rm str}(\zeta_\perp)-F^{\rm A,B}_{\rm str}(1)$ for each block along the DG-DD path, where red and blue denote A and B blocks, respectively, solid curves represent $\epsilon = 1$, and dashed curves represent $\epsilon = 3$. (C) Effective extensional strain elastic constant $C_{\rm ext}$ of DD for a range of A block fractions $f_{\rm A}$ and conformational asymmetry $\epsilon$, with the transition between stability and instability highlighted at $C_{\rm ext} = 0$.
    }
\end{figure}

To test the influence of packing frustration on the metastability of the DD phase, we consider the linear pathway connecting cubic DG and cubic DD for conformationally symmetric diblock melts from mSST for $f_A = 0.29$ in Fig.~\ref{fig:6}(A,top).  
The relative stretching free energy $\Delta F_{\rm str.}$ of each block is plotted along this pathway as function of a one-dimensional reaction coordinate $0\leq \zeta_\perp\leq1$ in Fig.~\ref{fig:6}(B).   
Both endpoints are local minima of the matrix B-block free energy, separated be a local maximum at intermediate node fusion, which we attribute to a preference for cubic structures with matrix geometries that are more symmetric and uniform. 
In contrast, the stretching free energy of tubular A-block increases monotonically from its minimum for cubic DG to its maximum for cubic DD, indicating that it serves as the thermodynamic driving force for fission of the tetravalent DD mesoatoms.  
Because the total SST free energy per chain is simply three times the sum of stretching free energy of A and B blocks~\cite{Matsen2002}, the strongly unfavorable entropic stretching cost of A-blocks in tetrahedral nodes overwhelms the stabilizing effect of the local minimum in B-block stretching free energy at $\zeta_\perp = 0$, leading to a barrier-free fission pathway.

These differences in stretching free energy suggest an alternative mechanism for metastabilizing DD deriving from {\it elastic asymmetry} between the tubular and matrix forming blocks~\cite{Dimitriyev2023}. 
In particular, increasing the relative stiffness of matrix blocks over tubular blocks via the {\it conformational asymmetry} $\epsilon = a_A/a_B$ can be expected to shift the relative magnitude of the matrix-block entropy, which has a tendency to stabilize tetravalent nodes, relative to the tubular-block tendency to destabilize them.  
We test this by computing the $\epsilon = 3$ DG-to-DD transformation free energy landscape, shown in Fig.~\ref{fig:6}(A,bottom), at composition $f_A = 0.55$, again focusing on conditions where lamellae and cylinders compete for equilibrium. 
In contrast to the conformationally-symmetric case, the $\epsilon=3$ landscape shows clear evidence of metastability of cubic DD, with evidence of a transition state at the saddle point intermediate to cubic DG and DD.
As shown in Fig.~\ref{fig:6}(B), we find that the stretching free energy of the A-block increases by a similar amount along the path for $\epsilon = 3$ as the $\epsilon = 1$ case, indicating that the increase in the fraction of A-block contributing to the stretching costs is balanced by its increased flexibility. 
In contrast, matrix B-block experiences a much deeper local minimum for $\epsilon = 3$ at cubic DD, sufficient to stabilize tetravalent nodes against fission.  

To test stability at {\it finite segregation} conditions, we determine $C_{\rm ext}$ over a wide window of compositions and elastic asymmetries $0.5 \leq \epsilon \leq 2.0$ with $\chi N = 25$ shown in Fig.~\ref{fig:6}(C) and more general $\chi N$ shown in SI Fig.S6.
We find that cubic DD is stabilized for asymmetry that stiffens the matrix B-block ($\epsilon \gtrsim 1.05$) and for sufficiently low A-fraction, further confirming the intuition that the instability of tetrahedral nodes is related to packing frustration of the tubular domain.

\section{Conclusions}

While DD has been widely assumed to be the closest competitor to DG for BCP melt assembly, under broad conditions, including the simplest case of conformationally symmetric diblocks, cubic DD is unstable via continuous transformation to cubic DG as well as metastable tetragonal double-network T\textsuperscript{111}.  This tendency for DD node splitting is notably consistent with a so-called ``liquid network" picture of double-networks, in which length-minimization of the skeletal graph serves as proxy for assembly free energy~\cite{Dimitriyev2024}.
We note further that there are additional continuous paths between DD and DG, including a rhombohedral distortion, which according to geometric measures of inhomogeneity in minimal surfaces could be potentially favorable over the tetragonal one~\cite{Schroder-Turk2006}.  
However, our SCFT results suggest that simple shear deformations of cubic DD increase the free energy so relaxation to cubic DG is more likely to occur along the tetragonal transformation path for BCP assemblies.


The predicted lack of metastability further confounds observations of DD morphologies in nominally single component, low conformational asymmetry diblocks~\cite{Chang2021,Feng2023,Shan2024}.
Given the relatively small fraction of ``guest'' molecules needed to confer metastability in DD (Fig.~\ref{fig:5}), it is possible even small amounts of contaminants or dispersity in BCP themselves could play an outsized role in shaping the free energy landscape of double-network structures.  
A further possibility is surface interactions at the sample boundary~\cite{Magruder_2024} could favor the high cubic symmetry of DD and promote its metastability, a scenario that might be particularly relevant to observations of DG-to-DD interconversion in microspheres of PS-PDMS block copolymers~\cite{Shan2024}.   
While in the present study we have considered only symmetry-breaking transformations that are themselves triply-periodic, we note that physical transformations between these networks involve even more complex, non-affine deformations, including spatial gradients of the local transformation pathway here.  
It remains to be understood how the gradients at the interphase boundary between DD and DG select the size of that gradient zone as well as relative orientations of cubic networks formed on either side of the boundary.

\begin{acknowledgement}
The authors gratefully acknowledge valuable discussions with E.~Thomas and thank M. Matsen for sharing data on SCFT phase boundaries. This research was supported by the U.S. Department of Energy (DOE), Office of Basic Energy Sciences, Division of Materials Sciences and Engineering, under award DE-SC0022229. SCFT calculations were performed on the UMass Unity Research Computing Platform at the Massachusetts Green High Performance Computing Center.

\end{acknowledgement}

\begin{suppinfo}




    



    

\end{suppinfo}

\bibliography{refs}

\end{document}



\title{Supplementary Information: \\Not even metastable: Cubic double-diamond in diblock copolymer melts}

\author{Micheal S. Dimitriyev}
\email{msdim@tamu.edu}
\affiliation{Department of Materials Science \& Engineering, Texas A \& M University, College Station, Texas}
\affiliation{Department of Polymer Science \& Engineering, University of Massachusetts, Amherst, MA}
\author{Benjamin R. Greenvall}
\affiliation{Department of Polymer Science \& Engineering, University of Massachusetts, Amherst, MA}
\author{Rejoy Matthew}
\affiliation{Department of Polymer Science \& Engineering, University of Massachusetts, Amherst, MA}
\author{Gregory M. Grason}
\email{grason@umass.edu}
\affiliation{Department of Polymer Science \& Engineering, University of Massachusetts, Amherst, MA}





\renewcommand\thefigure{S\arabic{figure}}
\renewcommand\thesection{S\arabic{section}}
\renewcommand\theequation{S\arabic{equation}}

\maketitle

\section{\label{app:mSST} Medial Strong Segregation Theory (mSST)}

\subsection{The SST free energy}
In the limit of infinite segregation strength ($\chi N \to \infty$), the free energy of a microphase separated block copolymer melt is given by
\begin{equation}
    \begin{split}
    F &= \gamma A + F_{\rm str}^{\rm A} + F_{\rm str}^{\rm B} \\
    F_{\rm str}^\alpha &= \frac{1}{2}\kappa_{\alpha}I_{\alpha}
    \end{split}\, ,
\end{equation}
where $A$ is the IMDS area and $F_{\rm str}^\alpha$ is the stretching free energy cost of block $\alpha$.
The form of the polymer stretching cost originates from parabolic brush theory (assuming that each block forms a strongly-stretched polymer brush) and the geometric factor $I_\alpha$ is given by
\begin{equation}\label{eq:second_moment}
    I_{\alpha} = \int_{V_{\alpha}}{\rm d}V\, z^2\, ,
\end{equation}
where $V_\alpha$ represents the domain occupied by block $\alpha$ and $z$ is the minimum distance an arbitrary point in the brush domain is from the nearest IMDS~\cite{Milner1988}.
The interfacial free energy density is given by
\begin{equation}
    \gamma = k_B T\rho \overline{a}\sqrt{\frac{\chi}{6}}\left(\frac{2}{3}\frac{\epsilon^{3/2} - \epsilon^{-3/2}}{\epsilon - \epsilon^{-1}}\right)
\end{equation}
where $\rho$ is the segment density, $\overline{a} \equiv \sqrt{a_{\rm A}{a_{\rm B}}}$ is the geometric mean of the statistical segment lengths $a_{\alpha}$, and the $\epsilon \equiv a_{\rm A}/a_{\rm B}$ is the conformational asymmetry~\cite{Helfand1975}.
The entropic stiffness parameters $\kappa_\alpha$ are given by
\begin{equation}
    \begin{split}
        \kappa_\alpha &= \frac{3\pi^2 \rho k_B T}{4 N^2 \overline{a}^2}\overline{\kappa}_\alpha \\
        \overline{\kappa}_\alpha &= \left\{\begin{array}{cc}
        \frac{n_{\rm A}}{f_{\rm A}^2\epsilon} & \alpha = {\rm A} \\
        \frac{n_{\rm B}\epsilon}{(1-f_{\rm A})^2} & \alpha = {\rm B} \\
        \end{array}
        \right.
    \end{split}
\end{equation}
where $N$ is the degree of polymerization and $n_\alpha$ is the number of arms of block alpha ($n_\alpha =1$ for linear polymers).

\subsection{Construction of generating surfaces from skeletal graphs}


The mSST method relies on the construction of candidate structures from which we can determine a free energy per chain.
The candidate structures are determined from a suitable set of ``generating surfaces'' that can be constructed from fixed skeletal graph structures.
While the ideal generating surfaces tend to resemble the IMDSs of the target structures, we emphasize that the generating surfaces are not the actual IMDSs but a starting point for constructing suitable terminal surfaces through a medial axis transformation of the generating surface.

\begin{figure}[t!]
    \centering
    \includegraphics[width=4in]{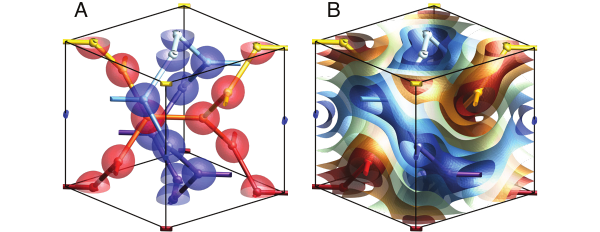}
    \caption{\label{fig:SI_levelsets} (A) Illustration of individual level sets generated by skeletal graph nodes (strut-generated level sets not shown). (B) Collection of level sets generated by sum over node and strut contributions.
    }
\end{figure}

Each fixed skeletal graph structure requires the construction of a pair of generating surfaces: one for each of the two network domains.
These surfaces are selected as the positive and negative level sets of a scalar function $\Psi(\mathbf{r})$ whose local maxima and minima lie on the skeletal graphs.
In order to construct $\Psi$ from the skeletal graphs, we draw on an electrostatic analogy wherein a fixed charge distribution determines a scalar field throughout all of space; in this way, the generating surfaces are specific equipotential surfaces of a given charge distribution, as shown in~\ref{fig:SI_levelsets}.
Each node is treated as a point source with a ``charge'' $q$ and point potential $\psi(\mathbf{r})$, as illustrated in Fig.~\ref{fig:SI_levelsets}(A); each strut is a linear charge distribution with linear ``charge density'' $\eta$.
Summing over all nodes and struts, we arrive at a potential of the form
\begin{equation}\begin{split}\label{eq:charge_sum}
    \Psi(\mathbf{r}) = &\sum_{m \in \Gamma_1}q_m\psi(\mathbf{r} - \mathbf{r}_m) + \sum_{\langle m,n\rangle \in \Gamma_1}\eta_{mn}\int_0^1{\rm d}t\,\psi\left(\mathbf{r} - (1-t)\mathbf{r}_m - t\mathbf{r}_n\right) \\ 
    &-\sum_{m \in \Gamma_2}q_m\psi(\mathbf{r} - \mathbf{r}_m) - \sum_{\langle m,n\rangle \in \Gamma_2}\eta_{mn}\int_0^1{\rm d}t\,\psi\left(\mathbf{r} - (1-t)\mathbf{r}_m - t\mathbf{r}_n\right)\, ,
\end{split}\end{equation}
where $\Gamma_1$ and $\Gamma_2$ denote the two graphs and $\mathbf{r}_{m,n}$ denote vertices in each graph; the notation $\langle m,n\rangle$ denote pairs of nodes that are joined by an edge.
Note that the charges on graph 2 have the equal and opposite charge of those on graph 1, allowing for each of the two network domains to be identified by the sign of $\Psi$, as shown in Fig.~\ref{fig:SI_levelsets}(B).

We find it convenient to represent the potential functions $\varphi$ using Gaussian distributions rather than standard electrostatic potentials; these are given by
\begin{equation}
    \psi(\mathbf{r} - \mathbf{r}_m) = \frac{1}{(2\pi w^2)^{3/2}}e^{-\frac{|\mathbf{r}-\mathbf{r}_m|^2}{2 w^2}}\, ,
\end{equation}
where the parameter $w$ fixes the width of the Gaussian; while this can be treated as a variational parameter, we have opted to fix $w = 0.1$.
While the integrals in~\ref{eq:charge_sum} can be exactly determined in terms of error functions, we found it sufficient and convenient to simply approximate $\int_0^1{\rm d}t\, \psi(t)$ as a sum $\delta t\sum_{i = 0}^{10}\psi(i\delta t)$ with $\delta t = 1/10$.
For each structure, we use the CGAL library to construct a pair of meshes for the two level sets.

Note that while the density fields formed by the melts have the symmetries of space groups $Ia\overline{3}d$ with nodes in the 16b Wyckoff positions (cubic DG), $Pn\overline{3}m$ with nodes in the 2a Wyckoff positions (cubic DD), and $P\overline{4}2m$ with nodes in the 2g/2h Wyckoff positions (T\textsuperscript{111}), the generating surface construction breaks the inversion symmetry that exchanges the two domains in each of these double networks.
The generating surface construction has space group symmetries $I4_132$ with nodes in the 8a/8b Wyckoff positions (cubic DG), $Fd\overline{3}m$ with nodes in the 8a/8b Wyckoff positions (cubic DD), and $I4_1/amd$ with nodes in the 8e Wyckoff positions (T\textsuperscript{111}).

\subsection{The medial construction of polymer packing environments}

Here we give a brief discussion of the medial SST approach; for a more detailed discussion, see~\cite{Reddy2022}.
Given a generating surface $\mathbf{G}$, we use the medial axis transformation to find suitable approximations for the terminal surfaces $\mathbf{T}_{\rm A}$ and $\mathbf{T}_{\rm B}$.
We use an algorithm that uses Voronoi tessellations (determined using the Voro++ package~\cite{Voro}) that extracts medial points from a set of generating points; see~\cite{Schroder2003}.
Since each point $\mathbf{p}_{\rm G}$ on the generating surface $\mathbf{G}$ can be mapped to two termini $\mathbf{p}_{\rm A} \in \mathbf{T}_{\rm A}$ and $\mathbf{p}_{\rm B} \in \mathbf{T}_{\rm B}$, there is a continuous family of points $\mathbf{p}(t) = (1-t)\mathbf{p}_{\rm A} + t \mathbf{p}_{\rm B}$ between the two termini.
In particular, there is a value of $t$ for which $\mathbf{p}(t)$ lies on the IMDS.
Each triangular mesh facet on the generating surface corresponds to triplets of points $\{\mathbf{p}_{\alpha,1}, \mathbf{p}_{\alpha,2},\mathbf{p}_{\alpha,3}\}$ on the two terminal surfaces.
The pair of triangles form the endcaps of a slender, wedge-like region of space representing a microscopic polymer packing environment.
We represent a points $\mathbf{X}$ in each wedge using barycentric coordinates,
\begin{equation}
    \mathbf{X}(u,v,t) = (1 - v)\mathbf{p}_{1}(t) + (1 - u)v\mathbf{p}_2(t) + uv\mathbf{p}_3(t)
\end{equation}
where $\mathbf{p}_i(t) = (1-t)\mathbf{p}_{i,{\rm A}} + t\mathbf{p}_{i,{\rm B}}$ and $u,v\in[0,1]$.
Since every point in space either lies on one of the terminal surfaces $\mathbf{T}_\alpha$ or between them, this allows us to define a \textit{terminal association map} that identifies any point in space to a pair of corresponding points on $\mathbf{T}_{\rm A}$ and $\mathbf{T}_{\rm B}$.

We divide each wedge (labeled $\mu$) into a pair of subwedges, corresponding to local A-block and B-block domains.
The subwedges are chosen through the placement of a dividing surface at a specific ``height'' $t^*$, such that $V_{\mu,{\rm A}}(t^*) = f_{\rm A}V_\mu$, where $V_{\mu}$ is the total volume of wedge $\mu$ and $V_{\mu,{\rm A}}$ is the volume of the subwedge filled with A-block material.
The collection of dividing surfaces amongst the wedges allows space to be partitioned into balanced A-block and B-block regions.
By tilting each surface such that a continuous surface is created, we arrive at a continuous IMDS.
Finally, we can calculate the integrals in Eq.~\ref{eq:second_moment} exactly, allowing for a calculation of the free energy per chain for each candidate structure.

Since these calculations are performed using arbitrary unit cell dimensions, we re-scale the dimensions of each candidate structure such that it contains an optimal number of chains.
This is done by re-scaling lattice parameters from the arbitrary computational scale $\tilde{D}$ to an equilibrium size scale $D = \sigma \tilde{D}$.
This adjusts the IMDS area $A \mapsto \sigma^2 \tilde{A}$, the volume $V \mapsto \sigma^3 \tilde{V}$, and the second moments of area $I_\alpha = \sigma^5 \tilde{I}_\alpha$.
This transforms the free energy per chain to
\begin{equation}
F_{\rm ch} = \frac{F}{n_{\rm ch}} = \sigma^{-1}\frac{\gamma N}{\rho}\frac{\tilde{A}}{\tilde{V}} + \frac{\sigma^2}{2}\sum_\alpha\frac{\kappa_\alpha N}{\rho}\frac{\tilde{I}_\alpha}{\tilde{V}}
\end{equation}
where the number of chains is given by $n_{\rm ch} = \rho V/N$.
Minimizing the above free energy with respect to $\sigma$ yields a scaling factor of the form
\begin{equation}
\sigma^* = \sqrt{\frac{2}{3\pi^{4/3}}}(\chi N)^{1/6}N^{1/2}\overline{a}\left(\frac{\tilde{A}}{\sum_\alpha \overline{\kappa}_\alpha \tilde{I}_{\alpha}}\right)^{1/3}
\end{equation}
so that the equilibrium free energy per chain is 
\begin{equation}
    F_{\rm ch} = \frac{3\pi^2}{4\tilde{V}}\left(\frac{2}{3}\frac{\epsilon^{3/2} - \epsilon^{-3/2}}{\epsilon - \epsilon^{-1}}\right)^{2/3}\left(\overline{\kappa}_{\rm A}\tilde{I}_{\rm A} + \overline{\kappa}_{\rm B}\tilde{I}_{\rm B}\right)^{1/3}\tilde{A}^{2/3}(\chi N)^{1/3}k_B T \, .
\end{equation}
Note that we recover the $(\chi N)^{1/3}k_B T$ scaling of the free energy per chain.

In order to minimize the free energy per chain, we use a Nelder-Meade algorithm to search over graph ``charge'' parameters, fixing the Gaussian widths $w = 0.1$.
The free energy landscapes are generated by minimizing the free energy and sweeping over tetragonal stretch parameter $\lambda$ by steps of $\Delta \lambda = 0.02$ and fusion parameters $\zeta_{\perp,\parallel}$ by steps of $\Delta \zeta = 0.05$.

\subsection{Fixed-volume mSST}

\begin{figure}[t!]
    \centering
    \includegraphics[width=4in]{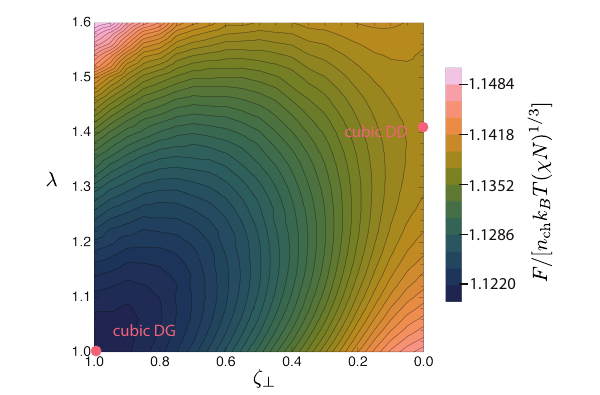}
    \caption{\label{fig:SI_fixedVolume_mSST} Free energy landscape determined by mSST under the constraint that the volume of each network is fixed to the volume of the cubic DD network.
    }
\end{figure}

To solve for equilibrium structures with fixed volume, we need to first determine the value of the volume that is being fixed.
Focusing on structures that have the volume of the cubic DD, we solve for the volume-optimized structure.
This amounts to determining the scaling factor $\sigma^*$ for the cubic DD structure.
Fixing the cubic DD as a ``reference structure'' with free energy $F_0$, the free energy of structures with the same volume is given by
\begin{equation}
    F = \frac{2}{3}F_0 \frac{\tilde{V}}{\tilde{V}_0}\left(\frac{\tilde{A}}{\tilde{A}_0} + \frac{1}{2}\frac{\overline{\kappa}_{\rm A}\tilde{I}_{\rm A}+\overline{\kappa}_{\rm B}\tilde{I}_{\rm B}}{\overline{\kappa}_{\rm A}\tilde{I}_{{\rm A},0}+\overline{\kappa}_{\rm B}\tilde{I}_{{\rm B},0}}\right)
\end{equation}
where $\tilde{A}_0$ is reference IMDS area, $\tilde{V}_0$ is the reference volume, and $\tilde{I}_{\alpha,0}$ are the reference second moments of volume for cubic DD.
Minimizing this form of the free energy with respect to the charge parameters, we arrive at the free energy landscape shown in Fig.~\ref{fig:SI_fixedVolume_mSST}.

\section{\label{app:tetr_SCFT}
Deformations in Self Consistent Field Theory}
\subsection{SCFT Workflow}
\begin{figure}[t!]
    \centering
    \includegraphics[width=6.5in]{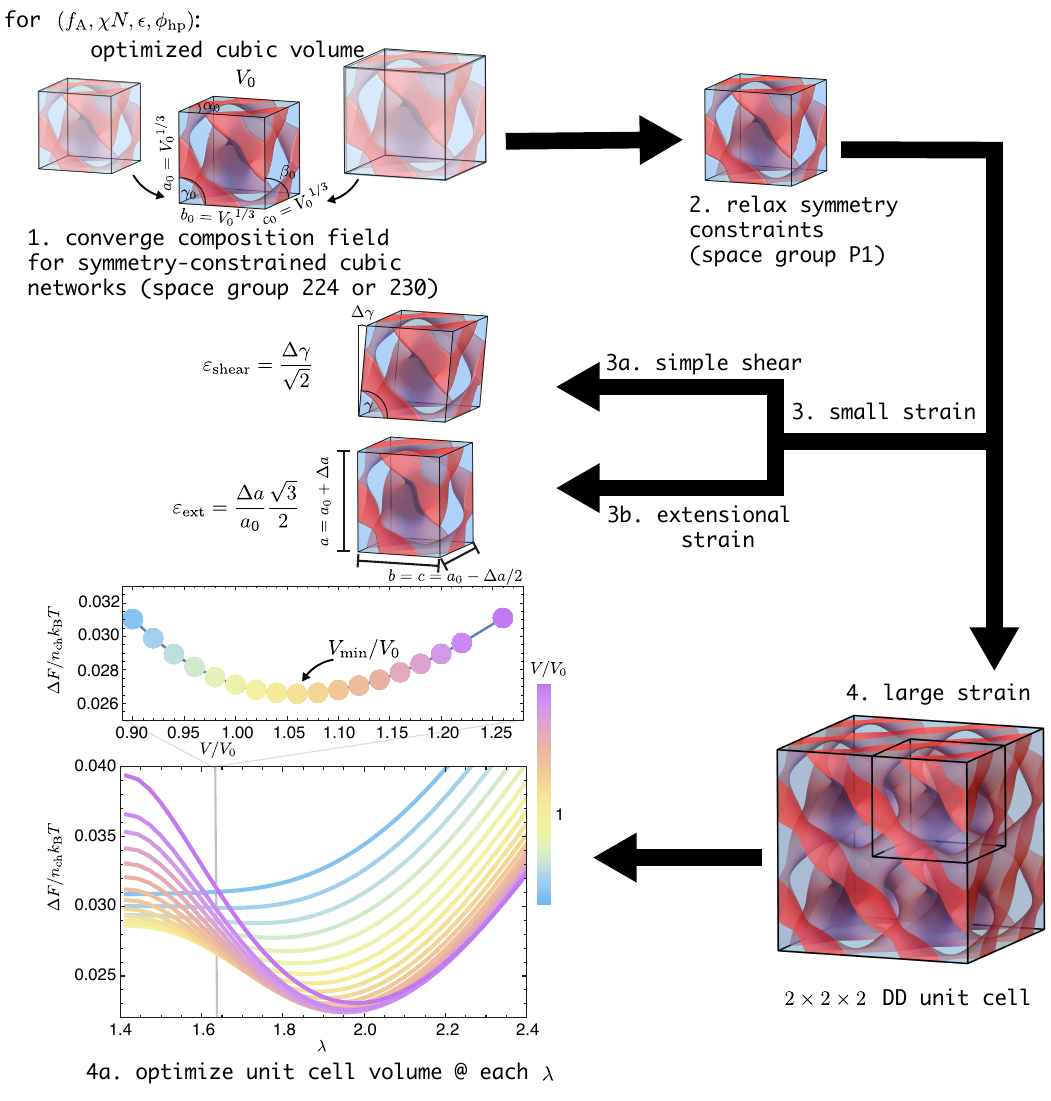}
    \caption{\label{fig:SI_fig_workflow}SCFT workflow for small strain (fixed unit cell volume) and tetragonal deformation pathway (large strain, variable unit cell volume).
    }
\end{figure}

To measure the thermodynamics of unit cell deformation in SCFT, we follow the procedure outlined in Fig.~\ref{fig:SI_fig_workflow}. 
First, we calculate the composition field and free energy per chain for a given set of chemical parameters ($f_{\rm A}, \chi N, \varepsilon, \phi_{\rm hp}$ for each network phase, assuming a cubic unit cell geometry with the appropriate constrained space group ($Pn\overline{3}m$ for DD and $Ia\overline{3}d$ for DG). Note that to probe the largest composition windows possible, we restrict the A-type polymer to always be the tubular domain, even under volume fraction conditions where inverse morphologies (tubes of B-type) would be favored (e.g. $f_{\rm A} > 0.5$ for $\epsilon = 1$).   
Our SCFT procedure is conducted using the open-source PSCF~\cite{Arora2016} software. 
This procedure additionally equilibrates the volume of the unit cell ($V_0$), such that the cubic unit cell lattice parameters are $a_0 = b_0 = c_0 = {V_0}^{1/3}$ and interaxial angles are $\alpha_0 = \beta_0 = \gamma_0 = \pi/2$. 
After obtaining a converged cubic field, we relax all symmetry constrains and further equilibrate the structure by specifying that the structure has space group $P1$ (using a triclinic unit cell initially maintaining the same lattice constants as the cubic cell). To probe stability at small strain, these triclinic fields are then be subject to either simple shear or extensional strain (like as shown in Fig.~4A), where (to linear order) the imposed strain of unit cell is given by


\begin{equation}\begin{split}
\varepsilon_{\mathrm{shear}} &\approx \frac{\Delta \gamma}{\sqrt{2}}\\
\varepsilon_{\mathrm{ext}} &\approx \frac{\Delta a}{a_0}\frac{\sqrt{3}}{2} 
\end{split}\end{equation}
where the lattice parameters for simple shear are given by
\begin{equation}
\begin{split}
a &= a_0 \\
b &= \frac{a_0}{1 + \sin \gamma} \\
c &=  a_0 \\
\alpha &= \beta = \frac{\pi}{2} \\
\gamma &= \frac{\pi}{2} + \Delta \gamma
\end{split}
\end{equation}
and the lattice parameters for extensional strain are given by
\begin{equation}
\begin{split}
a &= a_0 + \Delta a \\
b &= \frac{a_0}{\sqrt{1 + \Delta a/a_0}} \\
c &= \frac{a_0}{\sqrt{1 + \Delta a/a_0}} \\
\alpha &= \beta = \gamma = \frac{\pi}{2}
\end{split}
\end{equation}
Note that the above expressions for the lattice parameters are needed to ensure that the volume of the unit cell remains constant throughout each deformation path.

In practice, we deform the unit cell geometry (and calculate strain to quadratic order) following the procedure described in the Supporting Information of~\cite{Dimitriyev2024}. Due to the nature of the small deformations, the strain is conducted at the equilibrium cubic unit cell volume, $V_0$.  The calculated fields are also used to probe the consequences of larger tetragonal deformations (like as shown in Fig.~3). Here, we switch to using a larger DD unit cell, constructed of 8 copies of the original field ($2 \times 2 \times 2$ cells with cell parameter $a_0 = 2 {V_0}^{1/3}$), in order to capture larger symmetry-breaking related to the node fission. During these large deformations, it is no longer reasonable to assume the optimal field maintains the cubic volume, so we conduct tetragonal deformations with different values of a fixed unit cell volume, minimizing the volume for each value of $\lambda$ (see Fig.~\ref{fig:SI_fig_workflow}).

\subsection{SCFT convergence}

\begin{figure}[h!]
    \centering
    \includegraphics[width=6.5in]{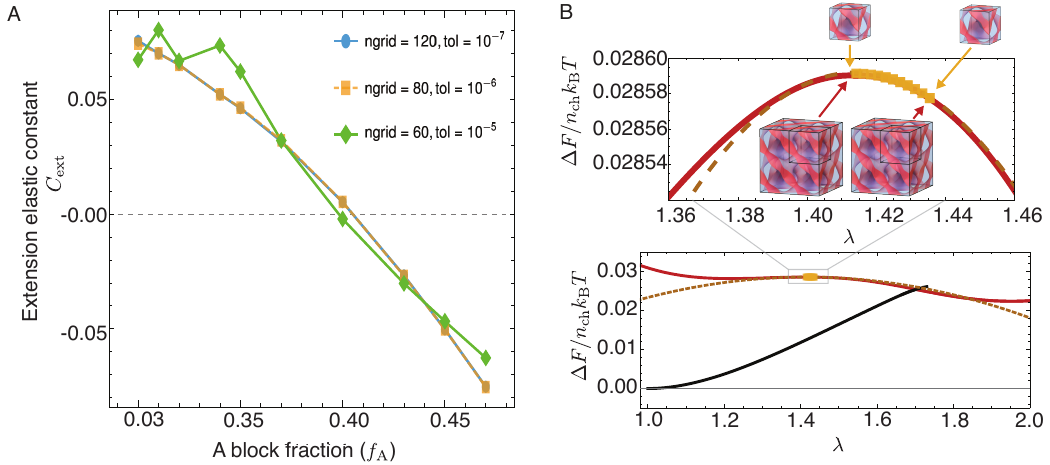}
    \caption{\label{fig:fig_SI_grid} Comparing the accuracy of SCFT parameters. (A) SCFT convergence of the cubic DD phase under uniaxial extension was evaluated by computing the extension elastic constant at  $\epsilon = 1.25$, using grid sizes $\texttt{ngrid} = 60, 80, 120$ and tolerances $10^{-5}, 10^{-6}, 10^{-7}$, respectively. Beyond a grid size of 80 with a tolerance of $10^{-6}$, the elastic constant shows no significant variation, indicating convergence.(B) Comparison of free energy accuracy for small strain ($1 \times 1 \times 1$ DD unit cell, yellow) and large strain ($2 \times 2 \times 2$ DD unit cell, red). The parabolic fit (dashed yellow) are to the small strain data; all curves agree in the region around cubic DD, $\lambda \simeq \sqrt{2}$. 
    }
\end{figure}
To evaluate the sensitivity of the computed elastic modulus under uniaxial extension to the SCFT numerical parameters of spatial grid resolution (\texttt{ngrid}) and self-consistency tolerance, we performed a convergence study at asymmetry factor $\epsilon=1.25$ and interaction strength $\chi N=25$. A sweep over the volume fraction $f_A$ was conducted using grid sizes of 60, 80, and 120 points, in combination with convergence tolerances(the \texttt{error\_max} parameter in the \texttt{iterate} section of PSCF param file ) of $10^{-5}, 10^{-6}$, and $10^{-7}$. For each parameter set, we calculated the free energy as a function of uniaxial strain and extracted the elastic modulus from the curvature of the resulting profile. We found that the curvature of the free energy and the corresponding modulus values exhibited negligible variation beyond $\texttt{ngrid} = 80$ and $\texttt{tol} = 10^{-6}$. As shown in Fig \ref{fig:fig_SI_grid}(A), the free energy–strain curves and the computed elastic moduli at $\texttt{ngrid} = 80, \texttt{tol} = 10^{-6}$ closely match those obtained at higher resolution ($\texttt{ngrid} = 120, \texttt{tol} = 10^{-7}$), indicating numerical convergence. Based on these results, we selected $\texttt{ngrid} = 80$ with a convergence tolerance of $10^{-6}$ for all subsequent small strain calculations, as this combination provides a practical compromise between computational efficiency and numerical accuracy. Due to larger differences in free energy at large strain, the tetragonal deformation sweeps (using the $2\times2\times2$ DD unit cell and primitive DG unit cell) are conducted using $\texttt{ngrid} = 80$ and a convergence tolerance of $10^{-5}$. However, we see good agreement between the $1\times1\times1$ unit cell resolved to $\texttt{tol} = 10^{-6}$ and $2\times2\times2$ unit cell resolved to $\texttt{tol} = 10^{-5}$ in the neighborhood around cubic DD (see Fig.~\ref{fig:fig_SI_grid}B. Therefore, our more detailed (small strain) approach shows utility in resolving finer features, however the coarser (large strain) approach is more than sufficient to capture larger thermodynamic features; together, they provide a comprehensive view of the local and far-field energetics.

\subsection{Skeletonization of SCFT Solutions}

To probe the non-affinity resulting from tetragonal deformation at finite segregation, we skeletonize the converged composition profiles generated via SCFT. Our approach is based on the approach developed in ref~\cite{Feng2019}, in which the composition field is sliced into planar sections, binarized based on composition to define the tubular domains, and then the centroids of each planar minority section are calculated. These centroids are then connected into 1D skeletal graph, where we require each node to have threefold functionality or higher (i.e. centroids with two or fewer neighbors are fused).

From the skeletal data, we classify each strut (given by the bond vector $\mathbf{\ell}$) as either participating in node fusion/fission (with strut length labeled $\ell_{\rm short}$ in Fig.~3) or not (with strut length labeled $\ell_{\rm long}$ in Fig.~3). For structures which are strained from cubic DG (e.g. $\zeta_\perp$), struts with $\ell_{\rm short}$ oriented orthogonal to the stretch direction (that is, $\mathbf{\ell} \cdot \hat{z} \simeq 0$) while structures strained from cubic DD (e.g. $\zeta_\parallel$) have their $\ell_{\rm short}$ struts aligned with their stretch direction ($|\mathbf{\ell} \cdot \hat{z}| \simeq 1$). Notably, we classify struts as either $\ell_{\rm short}$ or $\ell_{\rm long}$ irrespective of their relative length (for cubic DG, struts are all the same length, and extreme tetragonal stretches of cubic DD could yield struts with $\ell_{\rm short}$ > $\ell_{\rm long}$, though this does not occur under the deformation regimes considered here).

From the mSST skeletal graph construction, the strut length ratios $\ell_{\rm short}/ \ell_{\rm long}$ are fully encoded by the degrees of freedom ($\lambda$, $\zeta$):

\begin{equation}
    \begin{split}
\frac{\ell_{\rm short}}{\ell_{\rm long}} = -\frac{\sqrt{2}\zeta_\perp}{\sqrt{4 + 4 \zeta_\perp + \zeta_\perp^2 + \lambda^2}}\\
\frac{\ell_{\rm short}}{\ell_{\rm long}} = \frac{\lambda \zeta_\parallel}{\sqrt{4 + (1- \zeta_\parallel)^2\lambda^2}}
    \end{split}
\end{equation}

In SCFT, we instead solve these expressions for $\zeta_\perp$ and $\zeta_\parallel$, which allows for their calculation using the strut length ratio ($\ell_{\rm short}/ \ell_{\rm long}$) and tetragonal deformation ($\lambda$). 

\subsection{SCFT of Blends}
We study blends of AB diblock copolymer with A-type (minority) homopolymer in SCFT by ``labeling'' the homopolymer as a third chemical species in the PSCF software~\cite{Arora2016}. The diblock miscibility is given by the $N\chi_{\rm A_{\rm DBC}, B_{\rm DBC}}$, where $N$ is the diblock length and $\chi_{\rm A_{\rm DBC}, B_{\rm DBC}}$ is the Flory-Huggins interaction parameter (between diblock monomers of A- and B- type). We treat the A-type homopolymer as chemically equivalent to the A-type diblock (i.e. $\chi_{\rm A_{\rm DBC}, A_{\rm hp}} = 0$ and $\chi_{\rm A_{\rm DBC}, B_{\rm DBC}}= \chi_{\rm A_{\rm hp}, B_{\rm DBC}}$.

To determine the equilibrium (cubic) phase windows shown in Fig.~5(A), we determine the lowest free energy phase for each blend composition (i.e. value of $\phi_{\rm hp}$ out of a list of competing structures (BCC spheres, hexagonally-packed cylinders, lamella, as well as DD, DG, and Double Primitive networks). The narrow region of phase coexistence is determined via a common tangent construction. These same equilibrium windows were first reported in Fig.~8 of Ref.~\cite{Martinez-Veracoechea2009b}. 

\subsection{Additional SCFT Stability Analyses}
\begin{figure}[h]
    \centering
    \includegraphics[width=6.5in]{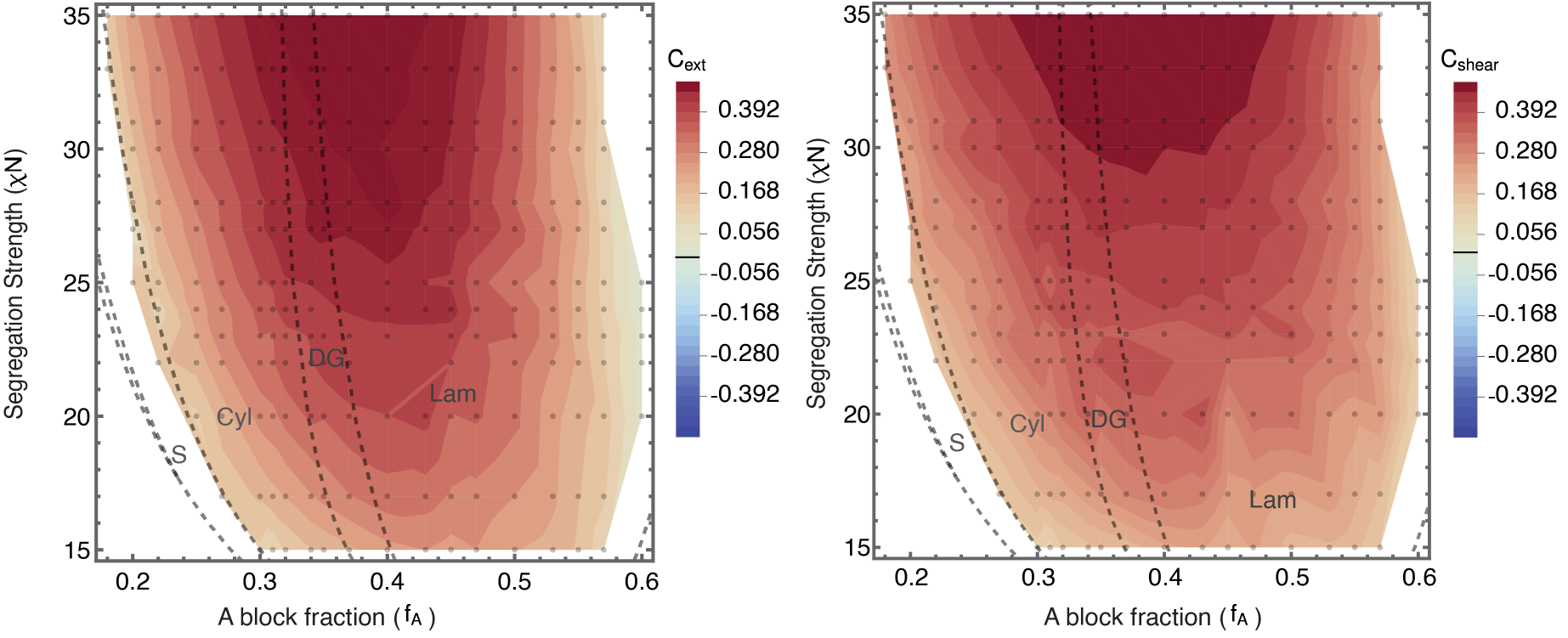}
    \caption{\label{fig:fig_SI_DG_ep_1}The stability of  conformally  symmetric DG as a function of $\chi N$ and A-block fraction $f_A$ are shown for  uniaxial extension and simple shear , superimposed on the diblock phase diagram from ~\cite{Matsen2012}. Red contours represent stable regions. Scale of concavity is shown in the colorbar, in units of $n_{\rm ch} k_B T$. Gray points correspond to SCFT data. 
    }
\end{figure}

In Figure~\ref{fig:fig_SI_DG_ep_1}, similar to Fig 4(C), we plot $C_{\rm shear}$ and $C_{\rm ext}$ for the cubic DG across a broad range of $\chi N$ and $f_A$ values for conformationally symmetric AB diblock copolymer melts. Both moduli remain positive ($C_{\rm shear},\, C_{\rm ext} > 0$) throughout the parameter space, indicating that DG is stable throughout this region.

\begin{figure}[h]
    \centering
    \includegraphics[width=6.5in]{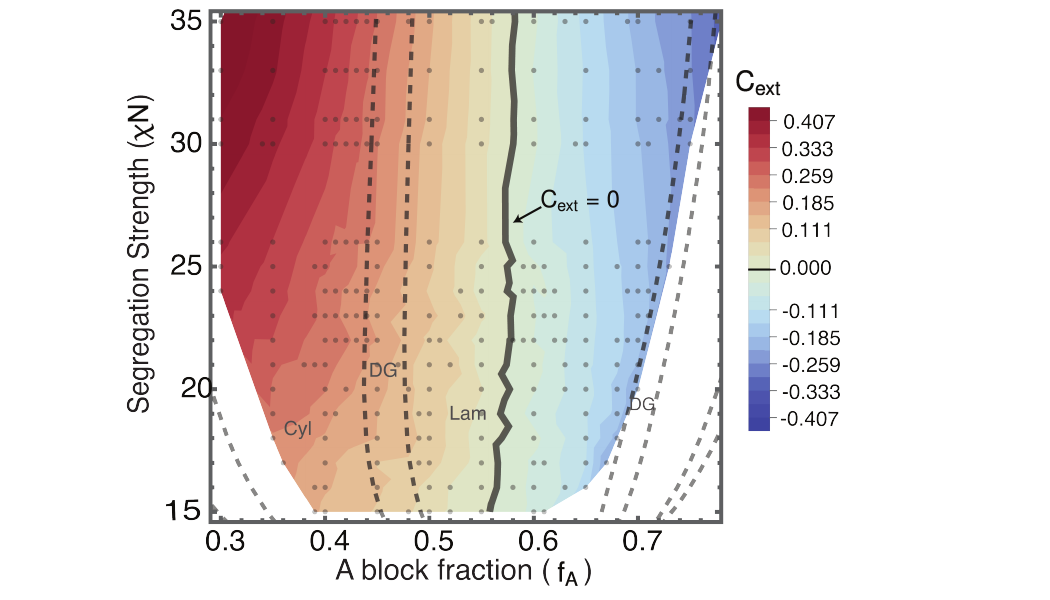}
\caption{\label{fig:fig_SI_DD_ep_2}Effective extensional elastic constant $C_{\rm ext}$ of the DD phase as a function of A-block fraction $f_{\rm A}$ and $\chi N$, for conformational asymmetry $\epsilon = 2$, superimposed on diblock phase diagram from \cite{Matsen1997}\cite{Xie2014} .The transition between stability and instability is indicated at $C_{\rm ext} = 0$.}
\end{figure}

In Figure~\ref{fig:fig_SI_DD_ep_2}, we determine $C_{\rm ext}$ over a wide window of $\chi N$ and $f_A$ at elastic asymmetry $\epsilon = 2$. The cubic DD phase is stable across the equilibrium DG window under extensional deformation across the entire $\chi N$ range, illustrating a large shift to lower composition for the boundary between instability/metastability when increasing from $\epsilon = 1$ to  $\epsilon = 2$.

\bibliography{refs}